\documentclass[preprintnumbers,superscriptaddress,pra,showkeys]{revtex4}
\usepackage{amsfonts}
\usepackage{amssymb}
\usepackage{amsmath}
\usepackage{epsfig}
\usepackage{graphicx}
\usepackage{color}

\begin{document}

\title{On relation between renormalized frequency and heat capacity for
particles in an anharmonic potential}
\author{Y. T. Liu}
\affiliation{School for Theoretical Physics, School of Physics and Electronics, Hunan
University, Changsha 410082, China}
\author{Y. H. Zhao}
\affiliation{School for Theoretical Physics, School of Physics and Electronics, Hunan
University, Changsha 410082, China}
\author{Y. Zhong}
\email{zhongy@hnu.edu.cn}
\affiliation{School for Theoretical Physics, School of Physics and Electronics, Hunan
University, Changsha 410082, China}
\author{J. M. Shen}
\email{shenjm@hnu.edu.cn}
\affiliation{School for Theoretical Physics, School of Physics and Electronics, Hunan
University, Changsha 410082, China}
\author{J. H. Zhang}
\affiliation{College of Material Sciences, Hunan University, Changsha 410082, China}
\author{Q. H. Liu}
\email{quanhuiliu@gmail.com}
\affiliation{School for Theoretical Physics, School of Physics and Electronics, Hunan
University, Changsha 410082, China}
\date{\today }

\begin{abstract}
For free particles in a simple harmonic potential plus a weak anharmonicity,
characterized by a set of anharmonic parameters, Newtonian mechanics asserts
that there is a renormalization of the natural frequency of the periodic
motion; and statistical mechanics claims that the anharmonicity causes a
correction to the heat capacity of an ideal gas in the anharmonic
potential.\ The orbital motion and thermal motion depend on the same
anharmonic parameters, but in different combinations. These two manners of
combinations are fundamentally different, demonstrating that statistical law
can not emerge from the many-body limit\emph{\ }of deterministic law for
one-body.
\end{abstract}

\author{}
\maketitle

\section{Introduction}

Anharmonicity plays a crucial role in modern physics, for instance,
classical anharmonic $\phi ^{4}$ model in quantum field theory, \cite{qft}\
and various anharmonic effects in condensed matter physics, \cite%
{solid1,solid2,solid3,solid4,solid5,solid6,solid7,solid8,solid9} and secular
evolution in planetary orbits. \cite{liu19} In the perturbation theory, the
anharmonicity is usually associated with the renormalization of the natural
frequency to remove the superficial divergence, i.e., to eliminate the
secular term in the naive solution; \cite{liu19,RONALD,PLRE,liu} and the
relationship between the renormalization in quantum field theory and removal
of the secular term in perturbation expansion has been a subject under
intensive investigations. \cite{QFT1,QFT2,QFT3,QFT4,QFT5,QFT6,QFT7,QFT8}
However, there are still problems not yet fully understood. On one hand, in
Newtonian mechanics the anharmonicity usually leads to a renormalization of
the natural frequency; and on the other, in statistical mechanics the
simplest situation is that the anharmonicity causes a correction to the heat
capacity for an ideal gas in the anharmonic potential. We have therefore two
different mechanics both start from the same Hamiltonian to deal with the
same many-body system: one is the Newtonian mechanics from which every
particle has its own orbit; and another is the statistical mechanics from
which every particle situates at a microstate with a definite probability.
An immediate question then arises: Can the statistical law emerge from the 
\emph{many-body limit }of deterministic law for one-body? This question may
be of \emph{fundamental importance}, and in present paper, an exactly
solvable one-dimensional system is used to understand this question in some
depth.

Assume that there is a particle of mass $m$ moving in a potential field $%
U\left( x\right) $ $\left( x\in \left( -\infty ,\infty \right) \right) $
given by,%
\begin{equation}
U\left( x\right) =U\left( 0\right) +\frac{1}{2}m\omega _{0}^{2}\ell
^{2}\left( \left( \frac{x}{\ell }\right) ^{2}-a\left( \frac{x}{\ell }\right)
^{3}+b\left( \frac{x}{\ell }\right) ^{4}+c\left( \frac{x}{\ell }\right)
^{5}+d\left( \frac{x}{\ell }\right) ^{6}\right) ,  \label{ur}
\end{equation}%
where $\omega _{0}$ is the natural frequency of the unperturbed potential $%
m\omega _{0}^{2}x^{2}/2$, and $a$, $b$, $c$, and $d$ are four small
dimensionless parameters accounting for various orders of the anharmonicity,
and $\ell \left( \neq 0\right) $ is a characteristic length accounting for,
e.g. the anharmonicity and we usually set $U\left( 0\right) =0$. We will
call these four constants $a$, $b$, $c$, and $d$ as anharmonic parameters.\
Once the anharmonicity happens at infinity $\ell \rightarrow \infty $, $%
U\left( x\right) $ reproduces the usual harmonic one. The characteristic
length $\ell $ can be conveniently chosen to be the amplitude of the initial
position, and can in fact be freely specified because our conclusion is
independent of its specific value. When $c=d=0$, the potential (\ref{ur})
becomes, 
\begin{equation}
U\left( x\right) =\frac{1}{2}m\omega _{0}^{2}\ell ^{2}\left( \left( \frac{x}{%
\ell }\right) ^{2}-a\left( \frac{x}{\ell }\right) ^{3}+b\left( \frac{x}{\ell 
}\right) ^{4}\right) .  \label{UX}
\end{equation}%
Landau used such a form of potential (\ref{UX}) describing the anharmonicity
of the vibrations and their interaction with the rotation within a diatomic
molecule (see Eq. (49.11) in Ref. \cite{Landau}), and we follow Landau's
convention \cite{Landau} to take a negative sign before first order
anharmonicity $a\left( x/\ell \right) ^{3}$ in (\ref{ur}) and (\ref{UX})
though the interval of $x$ in Landau model \cite{Landau} is half space $x\in
\left( 0,\infty \right) $ but what we are interested in is the full one.

Every particle in the potential $U\left( x\right) $ (\ref{ur}) moves along
an exclusive trajectory, no matter what energy it has. However, each
trajectory has its own frequency provided that it takes an exclusive value
of energy. As we show shortly (c.f. Eqs. (\ref{newt1}) and (\ref{renormf2}%
)), we can expand the renormalized frequency up to order $\left( x/\ell
\right) ^{4}$ in the following form, 
\begin{equation}
\omega \approx \omega _{0}\left( 1+\chi ^{\left( 1\right) }\mu +\chi
^{\left( 2\right) }\mu ^{2}+\chi ^{\left( 3\right) }\mu ^{3}+\chi ^{\left(
4\right) }\mu ^{4}\right) ,  \label{reff}
\end{equation}%
where parameter $\mu $ is dimensionless parameter, defined by,%
\begin{equation}
\mu \equiv \frac{A}{\ell }\succ 0,  \label{mu}
\end{equation}%
where $A\equiv x\left( t=0\right) \succ 0$ is initial position of the
particle, which can also be used to characterize the value of the energy the
particle takes, and once letting $\ell =A$, we have $\mu =1$. To note that
the parameter $\ell $ must be the same in both Newtonian mechanics and
statistical physics, otherwise the comparison of their results is
meaningless. The common features all trajectories in the potential (\ref{ur}%
) share are from (\ref{reff}) $\chi ^{\left( i\right) }=$ $\chi ^{\left(
i\right) }\left( a,b,c,d\right) $ ($i=1,2,3,4$), and we call $\chi ^{\left(
i\right) }$ the $i$-th order \emph{orbital anharmonicity} (OA). It is worth
stressing that OAs are independent of initial conditions $x$ and $dx/dt$,
whose possible uncertainty or stochasticity does not effect OAs. In
analogue, we will introduce $i$-th order \emph{thermal anharmonicity }$\zeta
^{\left( i\right) }=$ $\zeta ^{\left( i\right) }\left( a,b,c,d\right) $ in
the similar expansion of the heat capacity, 
\begin{equation}
C\approx C_{0}\left( 1+\alpha _{1}\zeta ^{\left( 1\right) }+\alpha
_{2}^{\left( 2\right) }\zeta ^{\left( 2\right) }+\alpha _{3}\zeta ^{\left(
3\right) }+\alpha _{4}\zeta ^{\left( 4\right) }\right) ,  \label{C}
\end{equation}%
where $\alpha _{i}$ are some expansion coefficients, and $C_{0}=Nk_{B}$ with 
$N$ the number of the particle and $k_{B}$ the Boltzmann constant. The key
finding of the present study is, 
\begin{equation}
\zeta ^{\left( i\right) }\text{ (}i=1,2,3,4\text{) is linearly independent
of OAs (}\chi ^{\left( 1\right) },\chi ^{\left( 2\right) },\chi ^{\left(
3\right) },\chi ^{14}\text{).}  \label{key}
\end{equation}

We are confident that it is a completely novel and physically significant
result for OA and \emph{thermal anharmonicity} offer proper and faithful
characterization of the anharmonicity from the point of Newtonian dynamics
and thermodynamics, respectively. The importance of the inequivalence
between OA and \emph{thermal anharmonicity} can be understood from the
opposite but untrue limit: If $\chi ^{\left( i\right) }=\zeta ^{\left(
i\right) }$, we could safely say that statistical law for the many-body
system and the many-body limit of Newtonian mechanics for every particle in
it are at least heavily overlapped and are even of same origin in nature.
Otherwise, Eq. (\ref{key}) strongly suggests that the statistical law can
not emerge from the many-body limit\emph{\ }of deterministic law for
one-body. In very rough terms, the molecular dynamics can not exactly and
completely reproduce all thermodynamic results. 

This paper is organized as follows. Section II and III give the detailed
steps of calculations of $\omega $ and $C$ with $c=d=0$, and Section IV
presents only the final results of both $\omega $ and $C$ with nonvanishing $%
c$ and $d$, and all calculational steps are omitted. Explicitly, in section
II, we utilize the Poincare--Lindstedt method to solve the equation of
motion of position $x$ in terms of time $t$, from which we see in detail how
the natural frequency is renormalized. In section III, the anharmonicity
induced correction of the heat capacity is calculated and an order-by-order
comparison between the renormalized frequency and the heat capacity is made.
In section IV, the potential containing higher order\ anharmonicities with $%
c\neq 0$ and $d\neq 0$ in (\ref{ur}) is studied, and we see clearly that
heat capacity $C$ not only involves these OAs $\chi ^{\left( 1\right) },\chi
^{\left( 2\right) },\chi ^{\left( 3\right) },$ and $\chi ^{\left( 4\right) }$
but also the anharmonic parameters $a,b,c$, and $d$. In final section V, a
brief conclusion is given.

\section{Renormalized frequencies for second order anharmonic oscillator}

The equation of motion for one particle in the potential (\ref{UX}) is,%
\begin{equation}
m\frac{d^{2}x}{dt^{2}}=-\frac{dU\left( x\right) }{dx}=-m\omega
_{0}^{2}\left( x-\frac{3}{2}a\ell \left( \frac{x}{\ell }\right) ^{2}+2b\ell
\left( \frac{x}{\ell }\right) ^{3}\right) ,  \label{eom}
\end{equation}%
where the initial conditions at instant $t=0$ are, 
\begin{equation}
x\left( 0\right) =A,\frac{dx(0)}{dt}=0.  \label{inits}
\end{equation}%
Making a variable transform, 
\begin{equation}
x(t)\rightarrow \ell \varphi (t),  \label{tranform}
\end{equation}%
we have from (\ref{mu}), (\ref{eom}) and (\ref{inits}), 
\begin{equation}
\frac{d^{2}\varphi }{dt^{2}}=-\omega _{0}^{2}\left( \varphi -\frac{3}{2}%
a\varphi ^{2}+2b\varphi ^{3}\right) ,\varphi \left( 0\right) =\mu ,\frac{%
d\varphi (0)}{dt}=0.
\end{equation}%
To this equation, no exact solution is possible due to the nonlinearity in $%
\varphi $, and even worse, the regular perturbation approaches fail for they
lead to the secular term in the solutions of $\varphi =\varphi (t)$.
Instead, the Poincare--Lindstedt method gives uniformly valid asymptotic
expansions for the periodic solutions of weakly nonlinear oscillations. \cite%
{RONALD,PLRE,liu} By the method, we mean that following three
transformations must be done simultaneously, 
\begin{subequations}
\begin{eqnarray}
\omega _{0} &\rightarrow &\omega =\omega _{0}+\omega _{1}+\omega _{2}+...,
\label{newt1} \\
t &\rightarrow &\tau =\frac{\omega _{0}}{\omega }t,\text{ }  \label{newt2} \\
x\left( t\right) &\rightarrow &\ell \xi \left( \tau \right) =x\left( t\left(
\tau \right) \right) ,  \label{newt3}
\end{eqnarray}%
where $\omega _{1}\sim $ $O(a)$ and $\omega _{2}\sim $ $O(a^{2})\sim O(b)$
are the first and second order renormalization of the frequency, and so
forth. In the same time, we have, 
\end{subequations}
\begin{equation}
\xi (\tau )\approx \xi _{0}(\tau )+\xi _{1}(\tau )+\xi _{2}(\tau )+...
\label{ALL}
\end{equation}%
in which $\xi _{0}(\tau )$ is the equation of motion for the unperturbed
oscillator satisfying the initial conditions,%
\begin{equation}
\xi _{0}(0)=\mu ,\frac{d\xi _{0}(0)}{d\tau }=0,
\end{equation}%
and $\xi _{1}\sim $ $O(a)$ and $\xi _{2}\sim $ $O(a^{2})$ are the first and
second order correction of the position $\xi (\tau )$, with the initial
conditions, respectively,%
\begin{equation}
\xi _{i}(0)=0,\frac{d\xi _{i}(0)}{d\tau }=0,(i=1,2).
\end{equation}

The correct equation of motion takes the following form, accurate up to $%
O(b) $ or $O(a^{2})$,%
\begin{equation}
\frac{d^{2}\xi }{d\tau ^{2}}\approx -\left( \omega _{0}+\omega _{1}+\omega
_{2}\right) ^{2}\left( \xi -\frac{3}{2}a\xi ^{2}+2b\xi ^{3}\right) ,
\label{meom}
\end{equation}%
The zeroth, first, and second order equations of motion of Eq. (\ref{meom})
are, respectively,%
\begin{gather}
\frac{d^{2}\xi _{0}}{d\tau ^{2}}+\omega _{0}^{2}\xi _{0}=0,  \label{zeroth}
\\
\frac{d^{2}\xi _{1}}{d\tau ^{2}}+\omega _{0}^{2}\xi _{1}+\left( -\frac{3}{2}%
a\omega _{0}^{2}\xi _{0}^{2}+2\omega _{1}\omega _{0}\xi _{0}\right) =0,
\label{firt} \\
\frac{d^{2}\xi _{2}}{d\tau ^{2}}+\omega _{0}^{2}\xi _{2}+\left( 2\omega
_{1}\omega _{0}-3a\omega _{0}^{2}\xi _{0}\right) \xi _{1}+2\omega
_{0}^{2}b\xi _{0}^{3}-3a\omega _{1}\omega _{0}\xi _{0}^{2}+\left( 2\omega
_{2}\omega _{0}+\omega _{1}^{2}\right) \xi _{0}=0.  \label{second}
\end{gather}%
The zeroth order equation of motion gives the usual harmonic oscillatory
solution,%
\begin{equation}
\xi _{0}(\tau )=\mu \cos \left( \omega _{0}\tau \right) .  \label{0th}
\end{equation}%
The naive solution of the first order equation of motion is then,%
\begin{equation}
\xi _{1}(\tau )=-\omega _{1}\tau \mu \sin \left( \omega _{0}\tau \right) +%
\frac{1}{4}a\mu ^{2}\left( 3-2\cos \left( \omega _{0}\tau \right) -\cos
\left( 2\omega _{0}\tau \right) \right) .
\end{equation}%
The first term in the right-hand side gives the divergent oscillatory
amplitude $\omega _{1}\tau \mu $ as time $\tau \rightarrow \infty $ with $%
\omega _{1}\neq 0$. To remove the divergence, we have to choose, 
\begin{equation}
\omega _{1}=0.
\end{equation}%
The correct first order solution of equation of motion is thus,%
\begin{equation}
\xi _{1}(\tau )=\frac{1}{4}a\mu ^{2}\left( 3-2\cos \left( \omega _{0}\tau
\right) -\cos \left( 2\omega _{0}\tau \right) \right) .  \label{1st}
\end{equation}%
The naive solution of the second order equation of motion is,%
\begin{eqnarray}
\xi _{2}(\tau ) &=&\frac{\mu \tau }{16}\left( 3\mu ^{2}\left(
5a^{2}-4b\right) \omega _{0}-16\omega _{2}\right) \sin (\omega _{0}\tau ) 
\notag \\
&&+\frac{\mu ^{3}}{16}\left( -12a^{2}+\left( \frac{29}{4}a^{2}-b\right) \cos
\left( \omega _{0}\tau \right) +4a^{2}\cos \left( 2\omega _{0}\tau \right)
+\left( b+\frac{3}{4}a^{2}\right) \cos \left( 3\omega _{0}\tau \right)
\right) .  \label{2nd}
\end{eqnarray}%
The first term in the right-hand side gives also the divergent oscillatory
amplitude as time $\tau \rightarrow \infty $, and this divergence can simply
be removed with $\omega _{2}$ being selected to satisfy, 
\begin{equation}
3\mu ^{2}\left( 5a^{2}-4b\right) \omega _{0}-16\omega _{2}=0.
\end{equation}%
We have the second order correction of the frequency $\omega _{2}$, 
\begin{equation}
\omega _{2}=\frac{3}{16}\left( 5a^{2}-4b\right) \mu ^{2}\omega _{0}=\chi
^{\left( 2\right) }\mu ^{2}\omega _{0}.
\end{equation}%
where $\chi ^{\left( 2\right) }$ is the second order OF\emph{\ }which is a
combination of second order anharmonic parameters $b$ and $a^{2}$, 
\begin{equation}
\chi ^{\left( 2\right) }\equiv \frac{3}{16}\left( 5a^{2}-4b\right) .
\end{equation}%
The correct second order solution of equation of motion (\ref{second}) is,%
\begin{equation}
\xi _{2}(\tau )=\frac{\mu ^{3}}{16}\left( -12a^{2}+\left( \frac{29}{4}%
a^{2}-b\right) \cos \left( \omega _{0}\tau \right) +4a^{2}\cos \left(
2\omega _{0}\tau \right) +\left( \frac{3}{4}a^{2}+b\right) \cos \left(
3\omega _{0}\tau \right) \right) .
\end{equation}

The important result is then that the natural frequency $\omega _{0}$ is
renormalized to be, up to accuracy of second order anharmonicity $%
O(a^{2})\sim O(b)$, 
\begin{equation}
\omega =\left( 1+\chi ^{\left( 2\right) }\mu ^{2}\right) \omega _{0}.
\label{renormf}
\end{equation}%
The oscillation is composed of a single prime frequency $\omega _{0}$ and
its higher order harmonics, 
\begin{eqnarray}
\xi (\tau ) &\approx &\mu \cos \left( \omega _{0}\tau \right) +\frac{1}{4}%
a\mu ^{2}\left( 3-2\cos \left( \omega _{0}\tau \right) -\cos \left( 2\omega
_{0}\tau \right) \right)  \notag \\
&&+\frac{\mu ^{3}}{16}\left( -12a^{2}+\left( \frac{29}{4}a^{2}-b\right) \cos
\left( \omega _{0}\tau \right) +4a^{2}\cos \left( 2\omega _{0}\tau \right)
+\left( b+\frac{3}{4}a^{2}\right) \cos \left( 3\omega _{0}\tau \right)
\right) .  \label{comb}
\end{eqnarray}

It is easily to verify that the energy is conserved for we have, 
\begin{equation}
E\left( t\right) =\frac{1}{2}m\left( \frac{dx}{dt}\right) ^{2}+U\left(
x\right) =\frac{1}{2}m\omega _{0}^{2}A^{2}\left( 1-a\mu +b\mu ^{2}\right)
=E\left( t=0\right) .
\end{equation}%
The anharmonicity induced correction of the energy is, 
\begin{equation}
\Delta E=\frac{1}{2}m\omega _{0}^{2}A^{2}\left( -a\mu +b\mu ^{2}\right) .
\end{equation}%
Requiring that the energy shift is small, we have, 
\begin{equation}
\frac{\left\vert \Delta E\right\vert }{\frac{1}{2}m\left( \omega
_{0}A\right) ^{2}}=\left\vert -a\mu +b\mu ^{2}\right\vert \ll 1.
\end{equation}%
The sufficient conditions for this equation are, 
\begin{equation}
\left\vert a\right\vert \ll 1,\left\vert b\right\vert \ll 1.  \label{ab}
\end{equation}%
This is what small constants $a$ and $b$ mean in Newtonian mechanics. Once
these conditions break, the perturbation method (\ref{meom}) does not apply.

\section{Second order anharmonicity induced correction of heat capacity}

For our purpose, we need to compute the partition function in Boltzmann
statistical mechanics, with $H=p^{2}/2m+U(x)$, 
\begin{equation}
Z\equiv \int_{-\infty }^{\infty }\int_{-\infty }^{\infty }\exp (-\beta H)%
\frac{dxdp}{h}=Z_{T}Z_{U},
\end{equation}%
where $h$ is the Planck's constant, and $Z_{T}$ is the momentum factor of
the partition function divided by $h$,%
\begin{equation}
Z_{T}\equiv \int_{-\infty }^{\infty }\exp (-\beta \frac{p^{2}}{2m})\frac{dp}{%
h}=\frac{\sqrt{2\pi }}{h}\sqrt{\frac{m}{\beta }},
\end{equation}%
and $Z_{U}$ is the configurational factor of the partition function, with
transform $x\rightarrow \ell \xi $, 
\begin{eqnarray}
Z_{U} &=&\int_{-\infty }^{\infty }\exp (-\beta U(x))dx  \notag \\
&=&\int_{-\infty }^{\infty }\exp \left( -\beta \left( \frac{1}{2}m\omega
_{0}^{2}\ell ^{2}\left( \left( \frac{x}{\ell }\right) ^{2}-a\left( \frac{x}{%
\ell }\right) ^{3}+b\left( \frac{x}{\ell }\right) ^{4}\right) \right)
\right) dx  \notag \\
&=&\ell \int_{-\infty }^{\infty }\exp \left( -\frac{\xi ^{2}}{2\eta ^{2}}%
\right) \exp \left( -\frac{-a\xi ^{3}+b\xi ^{4}}{2\eta ^{2}}\right) d\xi 
\notag \\
&\approx &\ell \int_{-\infty }^{\infty }\exp \left( -\frac{\xi ^{2}}{2\eta
^{2}}\right) \left( 1+\frac{a\xi ^{3}-b\xi ^{4}}{2\eta ^{2}}+\frac{1}{8}%
\left( \frac{a\xi ^{3}}{\eta ^{2}}\right) ^{2}\right) d\xi  \notag \\
&=&\sqrt{2\pi }\ell \eta \left( 1+\frac{3}{8}\left( 5a^{2}-4b\right) \eta
^{2}\right) ,  \label{partition}
\end{eqnarray}%
where $\eta $ is the dimensionless parameter, defined by,%
\begin{equation}
\eta \equiv \sqrt{\frac{1}{\beta m\omega _{0}^{2}\ell ^{2}}}=\sqrt{\frac{%
k_{B}T}{m\omega _{0}^{2}\ell ^{2}}}.  \label{at}
\end{equation}%
The partition function is then, 
\begin{equation}
Z\equiv Z_{T}Z_{U}\approx \frac{2\pi }{\beta h\omega _{0}}\left( 1+\frac{%
2\chi ^{\left( 2\right) }}{\beta m\omega _{0}^{2}\ell ^{2}}\right) =\frac{%
2\pi }{\beta h\omega _{0}}\left( 1+2\chi ^{\left( 2\right) }\eta ^{2}\right)
.  \label{partition2}
\end{equation}%
Once $\chi ^{\left( 2\right) }$ is negligible, the partition function
reduces to be,%
\begin{equation}
Z\approx \frac{2\pi }{\beta h\omega _{0}},
\end{equation}%
which leads to the energy equipartition result for the heat capacity of the
oscillatory degree of freedom, 
\begin{equation}
C_{0}=Nk_{B}.
\end{equation}%
The anharmonicity gives rise to the correction to the internal energy, up to
accuracy of second order anharmonicity $O(a^{2})\sim O(b)$, 
\begin{equation}
\Delta U^{(2)}=-N\frac{\partial }{\partial \beta }\ln \left( 1+\frac{2\chi
^{\left( 2\right) }}{\beta m\omega _{0}^{2}\ell ^{2}}\right) \approx -\frac{%
2\chi ^{\left( 2\right) }N}{m\omega _{0}^{2}\ell ^{2}}\frac{\partial }{%
\partial \beta }\frac{1}{\beta }=\frac{2\chi ^{\left( 2\right) }N}{m\omega
_{0}^{2}\ell ^{2}\beta ^{2}}.  \label{u-kai}
\end{equation}%
The corresponding correction to heat capacity is proportional to the first
power of the temperature via $\eta ^{2}$ (\ref{at}),%
\begin{equation}
\Delta C^{(2)}=\frac{\partial \Delta U}{\partial T}\approx 4\eta ^{2}\chi
^{\left( 2\right) }Nk_{B},  \label{c-kai}
\end{equation}%
which is also compatible with the Landau's result. \cite{Landau}

Evidently, in statistical mechanics, $\Delta C^{(2)}$ is the second order
quantity which linearly depends on $\chi ^{\left( 2\right) }$. Thus, we have
the heat capacity, 
\begin{equation}
C\approx \left( 1+4\eta ^{2}\chi ^{\left( 2\right) }\right) Nk_{B}.
\label{corr}
\end{equation}%
It takes the form (\ref{C}) which suggests in general a\ \emph{linear }%
dependence of $C$ on $\chi ^{\left( i\right) }$. Whether such a form (\ref{C}%
) persists for the heat capacity with higher order anharmonicities is an
interesting problem. In next section, we show that this \emph{linear }%
dependence on the OAs breaks.

\section{High order anharmonicities: renormalization of the natural
frequency and heat capacity correction}

When $c\neq 0$ and $d\neq 0$, the equation of motion (\ref{meom}) becomes, 
\begin{equation}
\frac{d^{2}\xi }{d\tau ^{2}}\approx -\left( \omega _{0}+\omega _{1}+\omega
_{2}+\omega _{3}+\omega _{4}\right) ^{2}\left( \xi -\frac{3}{2}a\xi
^{2}+2b\xi ^{3}+\frac{5}{2}c\xi ^{4}+3d\xi ^{5}\right) .  \label{newt}
\end{equation}%
Utilization of the Poincare--Lindstedt method to solve this equation of
motion of position $\xi $ in terms of time $\tau $, Eq. (\ref{newt}) gives
results for each order in the following. The first three solutions ($\xi
_{0}(\tau ),\xi _{1}(\tau ),\xi _{2}(\tau )$) are already given in (\ref%
{comb}), and the third order solution $\xi _{3}(\tau )$ is, 
\begin{equation}
\xi _{3}(\tau )=\frac{\eta ^{4}}{128}(\Lambda _{0}+\Lambda _{1}\cos (\omega
_{0}\tau )+\Lambda _{2}\cos (2\omega _{0}\tau )+\Lambda _{3}\cos (3\omega
_{0}\tau )+\Lambda _{4}\cos (4\omega _{0}\tau )),
\end{equation}%
where, 
\begin{subequations}
\begin{eqnarray}
\Lambda _{0} &=&3\left( 75a^{3}-84ab-40c\right) , \\
\Lambda _{1} &=&-119a^{3}+140ab+64c, \\
\Lambda _{2} &=&\frac{32}{3}\left( -9a^{3}+12ab+5c\right) , \\
\Lambda _{3} &=&-3a\left( 3a^{2}+4b\right) , \\
\Lambda _{4} &=&\frac{1}{3}\left( -3a^{3}-12ab+8c\right) .
\end{eqnarray}%
The fourth order solution $\xi _{4}(\tau )$ is, 
\end{subequations}
\begin{equation}
\xi _{4}(\tau )=\frac{3\eta ^{5}}{64}(\Omega _{0}+\Omega _{1}\cos (\omega
_{0}\tau )+\Omega _{2}\cos (2\omega _{0}\tau )+\Omega _{3}\cos (3\omega
_{0}\tau )+\Omega _{4}\cos (4\omega _{0}\tau )+\Omega _{5}\cos (5\omega
_{0}\tau )),
\end{equation}%
where, 
\begin{subequations}
\begin{eqnarray}
\Omega _{0} &=&-a\left( 75a^{3}-116ab-56c\right) , \\
\Omega _{1} &=&\frac{2357a^{4}}{64}+\frac{23b^{2}}{12}-\frac{1475a^{2}b}{24}-%
\frac{292ac}{9}-\frac{8d}{3}, \\
\Omega _{2} &=&\frac{16}{9}a\left( 18a^{3}-30ab-13c\right) , \\
\Omega _{3} &=&\frac{93a^{4}}{16}-2b^{2}-\frac{11a^{2}b}{4}+\frac{3ac}{4}+%
\frac{5d}{2}, \\
\Omega _{4} &=&\frac{1}{9}a\left( 3a^{3}+12ab-8c\right) , \\
\Omega _{5} &=&\frac{5a^{4}}{192}+\frac{b^{2}}{12}+\frac{5a^{2}b}{24}-\frac{%
11ac}{36}+\frac{d}{6}.
\end{eqnarray}%
The third order and fourth order normalized frequencies are, respectively, 
\end{subequations}
\begin{equation}
\text{$\omega $}_{3}=-a\chi ^{\left( 2\right) }\eta ^{3}\omega _{0}=\chi
^{\left( 3\right) }\eta ^{3}\omega _{0},
\end{equation}%
where $\chi ^{\left( 3\right) }$ is the third order OA, defined by,\emph{\ } 
\begin{equation}
\chi ^{\left( 3\right) }\equiv -a\chi ^{\left( 2\right) },
\end{equation}%
and,%
\begin{equation}
\text{$\omega $}_{4}=\frac{3\eta ^{4}}{1024}\left(
1155a^{4}-2200a^{2}b-1120ac+304b^{2}-320d\right) \omega _{0}=\chi ^{\left(
4\right) }\eta ^{4}\omega _{0},
\end{equation}%
where $\chi ^{\left( 4\right) }$ is the fourth order OA\emph{,} formed by a
non-trivial combination of all fourth order parameters $%
(a^{4},a^{2}b,ac,b^{2},d)$, defined by,%
\begin{equation}
\chi ^{\left( 4\right) }\equiv \frac{3\eta ^{4}}{1024}\left(
1155a^{4}-2200a^{2}b-1120ac+304b^{2}-320d\right) .
\end{equation}%
We anticipate that $\chi ^{\left( 3\right) }$ and $\chi ^{\left( 4\right) }$
will appear in the heat capacity.

To see how two quantities $\chi ^{\left( 3\right) }$ and $\chi ^{\left(
4\right) }$ may appear in the higher order corrections to heat capacity, let
us compute the partition function $Z\equiv Z_{T}Z_{U}$ with the full form of
potential (\ref{ur}). The result is, with calculations similar to (\ref%
{partition})-(\ref{partition2}), 
\begin{equation}
Z\equiv Z_{T}Z_{U}\approx \frac{2\pi }{\beta h\omega _{0}}\left( 1+2\chi
^{\left( 2\right) }\eta ^{2}+\gamma \eta ^{4}\right) .
\end{equation}%
where,%
\begin{eqnarray}
\gamma &=&\frac{15}{128}\left( 7\left( 33a^{4}-72a^{2}b-32ac+16b^{2}\right)
-64d\right)  \notag \\
&=&-8b\chi ^{\left( 2\right) }+8\chi ^{\left( 4\right) }.
\end{eqnarray}%
The anharmonicity gives rise to the correction to the internal energy in the
following, 
\begin{equation}
\Delta U=-N\frac{\partial }{\partial \beta }\ln \left( 1+\frac{2\chi
^{\left( 2\right) }}{\beta m\omega _{0}^{2}\ell ^{2}}+\frac{\gamma }{\left(
\beta m\omega _{0}^{2}\ell ^{2}\right) ^{2}}\right) \approx Nk_{B}T\left( 
\frac{2\chi ^{\left( 2\right) }}{\beta m\omega _{0}^{2}\ell ^{2}}+2\frac{%
\gamma -2\left( \chi ^{\left( 2\right) }\right) ^{2}}{\left( \beta m\omega
_{0}^{2}\ell ^{2}\right) ^{2}\beta }\right)
\end{equation}%
The corresponding correction to the heat capacity is,%
\begin{equation}
\Delta C=\frac{\partial \Delta U}{\partial T}=Nk_{B}\left( 4\chi ^{\left(
2\right) }\eta ^{2}-12\left( \left( \chi ^{\left( 2\right) }\right)
^{2}+4b\chi ^{\left( 2\right) }-4\chi ^{\left( 4\right) }\right) \eta
^{4}\right) .
\end{equation}%
Thus fourth order correction to the heat capacity does not depend on the OAs%
\emph{\ }alone because of presence of a term $b\chi ^{\left( 2\right) }$.
Collecting all results together, we have the renormalized frequency $\omega $
and heat capacity $C$, respectively, 
\begin{eqnarray}
\omega &\approx &\omega _{0}\left( 1+\chi ^{\left( 2\right) }\mu ^{2}-a\chi
^{\left( 2\right) }\mu ^{3}+\chi ^{\left( 4\right) }\mu ^{4}\right) ,\text{
and }  \label{renormf2} \\
C &\approx &Nk_{B}\left( 1+4\chi ^{\left( 2\right) }\eta ^{2}-12\left(
\left( \chi ^{\left( 2\right) }\right) ^{2}+4b\chi ^{\left( 2\right) }-4\chi
^{\left( 4\right) }\right) \eta ^{4}\right) .
\end{eqnarray}%
In terms of the thermal anharmonicity, we have two nonzero elements, the 2nd
and fourth order anharmonicity $\chi ^{\left( 2\right) }$ and $\left( \chi
^{\left( 2\right) }\right) ^{2}+4b\chi ^{\left( 2\right) }-4\chi ^{\left(
4\right) }$. It is evidently that both OA and the thermal anharmonicity have
only two common elements, the first and second order anharmonicities $\chi
^{\left( 1\right) }\left( =0\right) $ and $\chi ^{\left( 2\right) }$,
showing that from particle orbits statistical law can be largely
re-constructed. However, in essence, the thermal properties are independent
of the orbits for statistical mechanics dictate its own manner of dependence
on the anharmonic parameters.

Before closing this section, we discuss an interesting situation which a
little bit deviates the theme of present study. In statistical mechanics, we
can theoretically assume the characteristic length to be a "thermal one" $%
\ell =\sqrt{k_{B}T/m\omega _{0}^{2}}$. Then, we have $C=Nk_{B}$ which is
irrelevant to the anharmonicity. The demonstration is straightforward,
because the configurational factor of the partition function $Z_{U}$ is,
with transform $x\rightarrow \ell \xi $, 
\begin{eqnarray}
Z_{U} &=&\int_{-\infty }^{\infty }\exp (-\beta U(x))dx  \notag \\
&=&\int_{-\infty }^{\infty }\exp \left( -\frac{\beta }{2}m\omega
_{0}^{2}\ell ^{2}\left( \left( \frac{x}{\ell }\right)
^{2}+\sum_{j=3}a_{j}\left( \frac{x}{\ell }\right) ^{j}\right) \right) dx 
\notag \\
&=&\ell \int_{-\infty }^{\infty }\exp \left( -\frac{1}{2}\sum_{j=2}a_{j}\xi
^{j}\right) d\xi  \notag \\
&=&\ell f\left( \left\{ a_{j}\right\} \right) ,
\end{eqnarray}%
where $U(x)$ contains anharmonicity of arbitrarily high orders, and $a_{2}=1$
and $a_{j}$ $(j\geq 3)$ are anharmonic parameters, and $f\left( \left\{
a_{j}\right\} \right) \equiv \int_{-\infty }^{\infty }\exp \left( -\frac{1}{2%
}\sum_{j=2}a_{j}\xi ^{j}\right) d\xi $ is independent of the temperature.
Since the dependence of $Z_{U}$ on the temperature is via $\ell =\sqrt{%
k_{B}T/m\omega _{0}^{2}}$ only, the partition function becomes $Z=\left(
k_{B}T/\hbar \omega _{0}\right) f\left( \left\{ a_{j}\right\} \right) $. We
have immediately $C=Nk_{B}$. To note that the anharmonic potential of form (%
\ref{ur}) with $\ell =\sqrt{k_{B}T/m\omega _{0}^{2}}$ is in fact problematic
because by definition, Hamiltonian must be temperature-indepedent. However,
from the pure theoretical consideration, such a potential may be considered
as an effective one, which may enrich our understanding of the energy
equipartition theorem.

\section{Conclusions}

A particle in a simple harmonic potential is fully understood. However, once
the potential is added by some weakly anharmonic terms, the problem becomes
highly non-trivial as we learn from Kolmogorov--Arnold--Moser theorem and
Fermi--Pasta--Ulam--Tsingou nonlinear lattice oscillations. Once a particle
moves in the anharmonic potential field, the natural frequency must be
renormalized and the OAs can then be introduced; and the dependence of the
OAs on the anharmonic parameters is dictated by the Newtonian mechanics. For
a classical ideal gas in the same anharmonic potential field, the internal
energy and heat capacity have their own ways of dependence on the anharmonic
parameters, determined by the statistical mechanics, and the corresponding
so-called thermal anharmonicities are introduced. The OAs and thermal
anharmonicities reflect the anharmonicity in potential in mechanics and
thermodynamics, respectively. The first two order anharmonicities in the
mechanics and the thermodynamics are the same, whereas the third and fourth
order anharmonicities are different. Though no higher order anharmonicities
are calculated, we can conclude that the statistical law can not emerge from
the \emph{many-body limit }of deterministic law for few-body.

It seems to us that the clear difference between the OA and the thermal
anharmonicity is useful in exploring the relation between a single particle
that obeys the Newtonian mechanics and many particles that follows the
statistical mechanics. The applications to other problems are under
exploration.

\begin{acknowledgments}
QHL is grateful to the members, especially to Professor Hong Qian at
University of Washington and Hong Zhao at Xiamen University and Professor
Zhigang Zheng at Huaqiao University, of Online Club Nanothermodynamica
(Founded in June 2020) for extensive discussions of various problems in
statistical physics. We are indebted to Dr. Xinyuan Ai for participation of
the early stage of the present work. This work is financially supported by
National Natural Science Foundation of China under Grant No. 11675051 and
No. 11905056.
\end{acknowledgments}

\end{document}